\documentclass[%
aip,
amsmath,amssymb,
preprint,%
]{revtex4-2}

\usepackage{graphicx}
\usepackage{dcolumn}
\usepackage{bm}
\usepackage{physics}
\usepackage{color}
\usepackage[utf8]{inputenc}
\usepackage[T1]{fontenc}
\usepackage{mathptmx}
\usepackage{setspace}
\usepackage[margin=1in]{geometry}
\usepackage{physics}
\usepackage{color}
\usepackage{multirow}
\usepackage{float}
\usepackage{adjustbox}
\usepackage{booktabs, tabularx} 
\newcolumntype{C}{>{\centering\arraybackslash}X}
\usepackage[normalem]{ulem} 

\begin{document}
\title {High-sensitivity Fluorescence-Detected Multidimensional Electronic Spectroscopy Through Continuous Pump-probe Delay Scan}

\author{Amitav Sahu}
\affiliation{ 
	Solid State and Structural Chemistry Unit, Indian Institute of Science, Bangalore, Karnataka 560012, India
}%
\author{Vivek N. Bhat}
\affiliation{ 
	Solid State and Structural Chemistry Unit, Indian Institute of Science, Bangalore, Karnataka 560012, India
}%
\author{Sanjoy Patra}
\affiliation{ 
	Solid State and Structural Chemistry Unit, Indian Institute of Science, Bangalore, Karnataka 560012, India
}%
\author{Vivek Tiwari}%
\email{Author to whom correspondence should be addressed: vivektiwari@iisc.ac.in}
\affiliation{ 
	Solid State and Structural Chemistry Unit, Indian Institute of Science, Bangalore, Karnataka 560012, India
}%

\date{\today}

\begin{abstract}
Background-free fluorescence detection in multidimensional electronic spectroscopy promises high sensitivity compared to conventional approaches. Here we explore the sensitivity limits of multidimensional electronic spectroscopy. We present a fluorescence-detected multidimensional electronic spectrometer based on a visible white-light continuum. As a demonstration of sensitivity, we report room temperature two-dimensional coherence maps of vibrational quantum coherences in a laser dye at optical densities $\sim$2-3 orders of magnitude lower than conventional approaches. This high sensitivity is enabled by a combination of biased sampling along the optical coherence time axes and a rapid scan of the waiting time $T$ dimension at each time step.  A combination of acousto-optic phase modulation and phase-sensitive lock-in detection enables simultaneous collection of rephasing and non-rephasing signals and measurements of room temperature vibrational wavepackets even at the lowest ODs. Alternative faster data collection schemes, enabled by the flexibility of continuous pump-probe scanning approach, are also demonstrated.
\end{abstract}
\maketitle 

\section{Introduction}
Quantum relaxation in the condensed phase can involve numerous optical resonances coupled through the motions of electrons and nuclei. Spectral congestion can limit mechanistic insights into fundamental photophysical processes such as ultrafast energy and charge transfer reactions. Multidimensional electronic spectroscopy (MES) borrows concepts\cite{Jonas2003a} from Fourier transform multidimensional nuclear magnetic resonance (NMR) experiments\cite{Ernst1976} to spectrally decongest femtosecond relaxation dynamics into an evolving two-dimensional (2D) contour map along the excitation and detection frequency axes. The resulting 2D spectrum can provide frequency resolution limited only by the system. Evolving contour map snapshots along the pump-probe waiting time $T$, with time resolution limited by the excitation pulses, are reporters of electronic relaxation, as well as fluctuating nuclear environments and molecular structures, which are frozen during a femtosecond excitation. MES has advanced understanding of novel ultrafast processes by disentangling congested dynamics, such as those of polariton, plasmon, and molecular aggregate states in `plexitons'\cite{ZigmantasPlexiton2021} to coherence, energy, or charge transfer in photosynthetic proteins and whole cells\cite{Miller2020}. 

Interferometric detection of a phase-matched signal has sensitivity limitations for scatter prone samples such as whole cells\cite{Dahlberg2013,Dostal2016}. Improvements in experimental techniques\cite{OgilvieARPC} such as double chopping and spatially-encoded time delays have enabled scatter suppression to a certain extent. Action-based MES\cite{Tiwari2021} is an alternative approach to MES which promises high-sensitivity. For example, recent demonstrations\cite{Roeding2018, Bruder2018} of action-based MES through high-sensitivity detection of photoionization signals have opened a new paradigm in gas phase spectroscopy, and motivate similar explorations in fluorescence-based MES (fMES). fMES relies on detection of fluorescence arising from non-linear wavepacket interferometry\cite{CinaARPC}, to provide facile scatter suppression through optical filtering. 

Demonstrations of fully collinear fMES through 125-step shot-to-shot phase cycling\cite{Mueller2019} have also led to the isolation of higher order non-linear signals. Acousto-optic phase modulation (AOPM) based dynamic phase cycling\cite{Tekavec2007}, or spatial light modulator (SLM) based 27-step phase cycling approaches to fMES are well suited for higher repetition rate broadband light sources, and are more desirable for microscopy applications. Both approaches have demonstrated 2D coherence map measurements, which could be challenging to measure at room temperature, and also paved the way for spatially resolved fMES in a confocal geometry\cite{Goetz2018, Tiwari2018a}. The sensitivity of background-free fluorescence detection to measurements of weak nonlinear signals such as vibrational coherences at room temperature is expected to be high but not yet explored in either case. Parallels in 2DIR have already demonstrated\cite{tokma2018} feasible measurements of vibrational couplings in a 2DIR spectrum at concentrations $\sim$2x lower than typical upon fluorescence-encoding of vibrational excitations. 

This manuscript explores the sensitivity limits of the AOPM-based dynamic phase cycling approach to fMES. One way to improve sensitivity is through reductions in data collection time, such as the recent demonstration\cite{Ogilvie2021} of rephasing and non-rephasing 2D spectra at a single waiting time $T$ through rapid stage scan along the optical coherence axes. This demonstration is especially relevant in the context of spatially-resolved measurements when all other fMES approaches reported\cite{Tiwari2021} so far have relied on two orders of magnitude higher repetition rates. Notably, a combination of white-light continuum (WLC) based experiments and lower repetition rates makes such an approach highly desirable but equally more challenging. Here we demonstrate $\sim2-3$ orders of magnitude higher sensitivity in fMES than conventional MES by considering an alternative rapid scan approach.

Rapid scan pump-probe experiments and theoretical modeling from Moon et al.\cite{Moon1993} have suggested that rapid-scan approach leads to higher signal-to-noise (SNR) for high signal levels. Counterintuitively, in case of weak signals with dominant probe-induced noise, higher SNR can be achieved with a fast lock-in time constant compared to rapid scanning (see Figures 3 and 4 of ref. \cite{Moon1993}). This is made possible by lock-in demodulation minimizing the probe spectral noise density transferred to the signal. Further, typically all fMES approaches so far have relied on scanning the delay axis grid uniformly even though noise fluctuations in data points close to zero delay affect the signal the most.  Motivated by these considerations, we report a visible WLC based fMES approach that relies on biased stepwise sampling along the optical coherence axes, and continuous scanning of pump-probe delay $T$ instead. The advantages of phase-sensitive lock-in detection in the AOPM approach are maintained because the $T$ delay is rapidly scanned using a fast lock-in time constant. Continuous scan, that is, fine $T$ sampling, implemented here is relevant for detecting weak coherences along $T$ in WLC based MES setups, where light source stability across the entire bandwidth is typically lower\cite{Chergui2012,Lang2018} than that achievable through optical parametric amplification. Measurements with fine sampling along $T$ are more robust to laser power drifts with $1/\sqrt{N}$ improvements, where $N$ is the number of fine-sampled $T$ points. Further, the ability to rapid scan $T$ allows for more number of averages along $T$ necessary to break spectral and temporal correlations across the WLC bandwidth\cite{Bradler2014}.  Crucially, the stepwise sampling along optical coherence axes is \textit{biased} towards sampling uniform signal contours rather than grids. Additionally, continuous $T$ scanning per grid point allows us to freely optimize the time window over which the sample is continuously exposed to light. Together these innovations allow for unprecedented sensitivity improvements with detection of room temperature coherence maps in a laser dye at ODs $\sim$2-3 orders of magnitude lower than conventional MES approaches, along with signal isolation based on $T$ quantum beat phase. Our approach of rapid scan along $T$ and biased stepwise sampling along optical coherence axes harnesses the full potential of the AOPM approach through physical undersampling of optical coherences, is amenable to non-uniform sampling and signal reconstruction through compressive sensing algorithms\cite{Sanders2012}, and relevant for measurements at lower repetition rates or with low count rate sources such as entangled photon pairs\cite{Lavoie2020}. Substantial reduction in the sample exposure window also makes continuous $T$ scanning approach highly desirable for spatially resolved measurements.

\section{Experiment and Methods}
This section briefly discusses the theoretical foundation for the AOPM approach to fMES before describing our experimental approach. This is followed by a description of the data collection and analysis procedure.
\subsection{fMES Signal and Acousto-Optic Phase Modulation (AOPM)}\label{aopm}
Cina and Marcus have detailed the formalism\cite{CinaARPC,Tekavec2007} for fMES in terms of non-linear wavepacket interferometry (WPI) experiments, originally pioneered by Scherer et.al.\cite{Scherer1991}. This formalism will be described briefly in the context of the AOPM approach implemented here.

Consider a two-level system with state kets $\ket{g}$, $\ket{e}$ representing the two electronic states. The initial state at $t=0$ is $\ket{g}$. A first-order light-matter interaction at time $t_1$ results in the time-dependent state, $\ket{\psi^{(1)}(t)}=\frac{-i}{\hbar}{\int_{0}^{t}e^{i\omega_e(t-t_1)}(-\hat{\mu}_{eg}\cdot\vec{E}(t_1))e^{-i\omega_g t_1} d t_1}$ ,where, $\hat{\mu}_{ge}$ is the transition dipole vector from ground state $\ket{g}$ to excited state $\ket{e}$, and $\omega_g$, $\omega_e$ are the energy of the ground and excited state respectively. Converting this above equation into a Fourier transform\cite{JonasARPC2003}, the probability amplitude transfered to the excited state after a first-order light-matter interaction is $\frac{i}{\hbar}\hat{\mu}_{eg}\cdot \vec{E}(\omega_{eg})$. In a WPI experiment, the probability amplitude arising from different light-matter interactions interfere\sout{d} to result in modulations of excited state population, recorded as modulating fluorescence signals. For example, wavepacket interferences between first-order wavepackets such as  $\braket{\psi^{(1)}_{1}}{\psi^{(1)}_{2}}$ leads to fluorescence signals that are linear with excitation intensity, and depend on the relative delay between the two first-order interactions (at times $t_1$ and $t_2$ for $\vec{E}_1$ and $\vec{E}_2$, respectively). Note that parenthesis superscript denotes order of interaction and subscript denotes electric field labels. Only interference between wavepackets arising from \textit{different} electric fields depend on the experimentally controllable time delay between their envelopes. In the same fashion, the non-linear wavepacket overlaps in fMES arise from interferences\cite{Tekavec2007} such as $\braket{\psi^{(3)}_{124}}{\psi^{(1)}_{3}}$ and $\braket{\psi^{(3)}_{134}}{\psi^{(1)}_{2}}$ for ground state bleach (GSB) and excited state emission (ESE) signals, respectively.

The total electric field at a time $t_1$ can be written in the frequency domain as, $\vec{E}_i(\omega) = e_i(\omega) e^{i\phi_i(\omega)}e^{i\omega t_1}$, where the last phase term arises due to a shift $t_1$ in the time domain. $e_i(\omega)$ is the real electric field envelope. The spectral phase is given by $\phi_i(\omega)$. The orientationally averaged fluorescence signal $\braket{\psi^{(1)}_{1}}{\psi^{(1)}_{2}}$ can then be expressed as,
\begin{equation}
\braket{\psi^{(1)}_{1}}{\psi^{(1)}_{2}} =\frac{1}{3\hbar^2} e_1(\omega_{eg})e_2(\omega_{eg})\abs{\hat{\mu}_{ge}}^2 e^{i\Delta\phi_{21}(\omega_{eg})}e^{i\omega_{eg} t_{21}}
 \label{eq1} 
\end{equation}
It is assumed that the signal is time-integrated by a slow detector. In the semi-classical formulation of Heller and co-workers\cite{Heller1981}, the above wave packet overlaps can be interpreted as an autocorrelation function of the excited state wavepacket $\ket{\psi^{(1)}(t)}$, such that Fourier transform of this autocorrelation function can be directly related to the absorption cross-section spectrally filtered by the excitation electric fields. Just like autocorrelation between first-order wavepacket depends on delay $t_{21}$, overlaps between non-linear wavepackets such as $\braket{\psi^{(3)}_{124}}{\psi^{(1)}_{3}}$ and $\braket{\psi^{(3)}_{134}}{\psi^{(1)}_{2}}$ are three-point autocorrelation functions and depend on time delays $t_{21}$, $t_{32}$, $t_{43}$.

Eqn.\ref{eq1} indicates that an optical frequency $\omega_{eg}$ is directly sampled as the time delay $t_{21}$ is scanned. Further, a spectrally imbalanced interferometer, for example, due to spatial chirp in the acousto-optic modulator (AOM) diffracted pulse, imparts an extra phase $e^{i\Delta\phi_{21}(\omega)}$ which can distort real absorptive spectral lineshapes by mixing real and imaginary parts of the resulting spectra. In the AOPM approach, each pulse $1-4$ is phase modulated by a radio frequency phase, $\vec{E}_i(\omega) = e_i(\omega)e^{i\omega t_i}e^{i\phi_i(\omega)}c^{i\Omega_i.nT_R}$, where $T_R$ is the laser repetition rate and $n$ denotes the $n^{th}$ lasershot. Thus, the carrier-envelope phase of each arm is dynamically phase cycled at a unique rate determined by the set radio frequency $\Omega_i$. Taking this into account, the linear fluorescence signal becomes --
\begin{equation}
\braket{\psi^{(1)}_{1}}{\psi^{(1)}_{2}} =\frac{1}{3\hbar^2} e_1(\omega_{eg})e_2(\omega_{eg})\abs{\hat{\mu}_{ge}}^2 e^{i\Delta\phi_{21}(\omega_{eg})}e^{i\omega_{eg} t_{21}}e^{i\Omega_{21}.nT_R}
 \label{eq2} 
\end{equation}
with phase modulation in individual arms leading to an amplitude modulated signal at frequency $\Omega_{21}$. This modulation frequency uniquely distinguishes linear fluorescence signal arising from two diffreent electric field interactions, against those that arise from the same electric field. Additionally, a reference signal is generated by passing the interfering electric fields through a monochromator set at a frequency $\omega_M$. The resulting field autocorrelation signal is given by,
\begin{equation}
R(\omega_M,t_{21}) = e_{R1}(\omega_M)e_{R2}(\omega_M)e^{i\Delta\phi_{R21}(\omega_M)}e^{i\omega_Mt_{21}}e^{i\Omega_{21}.nT_R}
 \label{eq3} 
\end{equation}
The spectral phase difference $\Delta\phi_{R21}(\omega_M)$ between the reference electric fields after passing through the monochromator can be assumed to be negligible. The AOPM approach relies on detecting the wavepacket autocorrelation signal (Eqn.\ref{eq2}) relative to the optically generated reference (Eqn.\ref{eq3}). The resulting demodulated lock-in signal including the in-phase and in-quadrature components now modulates as $Z(t_{21}) \sim e^{i(\omega_{eg}-\omega_M)t_{21}}e^{i\Delta\phi_{21}(\omega_{eg})}$. Note that imbalanced spectral phase in the interfering electric fields will mix the absorptive and dispersive lineshapes resulting from Fourier transform along $t_{21}$. Crucially, the optical frequency $\omega_{eg}$ is now undersampled at $(\omega_{eg}-\omega_M)$ imparting passive phase stability against fluctuations $\delta t_{21}$ in delay line.
In case of non-linear WPI, four-wavemixing signal pathways, such as rephasing and non-rephasing signals, can be isolated based on unique radio frequency combinations, $\Omega_{-} = (-\Omega_1+\Omega_2) + (\Omega_3-\Omega_4)$ and $\Omega_{+} = (\Omega_1-\Omega_2) + (\Omega_3-\Omega_4)$, for rephasing and non-rephasing signals respectively. Similar to physical undersampling of linear signals, optical coherences along the first and third time intervals, $t_{21}$ and $t_{43}$, are undersampled as $(\omega_{eg}-\omega_{R1})$ and $(\omega_{eg}-\omega_{R2})$ respectively, where $\omega_{R1,2}$ are reference optical frequencies set by the repective interferometer. Note that, the four wavemixing signal lineshapes can also be distorted due to imbalanced spectral phase $\mp \Delta\phi_{21}(\omega_{eg})+\Delta\phi_{34}(\omega_{eg})$.

\subsection{Experimental Setup}\label{setup}
\begin{figure*}[!ht]
	\centering
	\includegraphics[width=5 in]{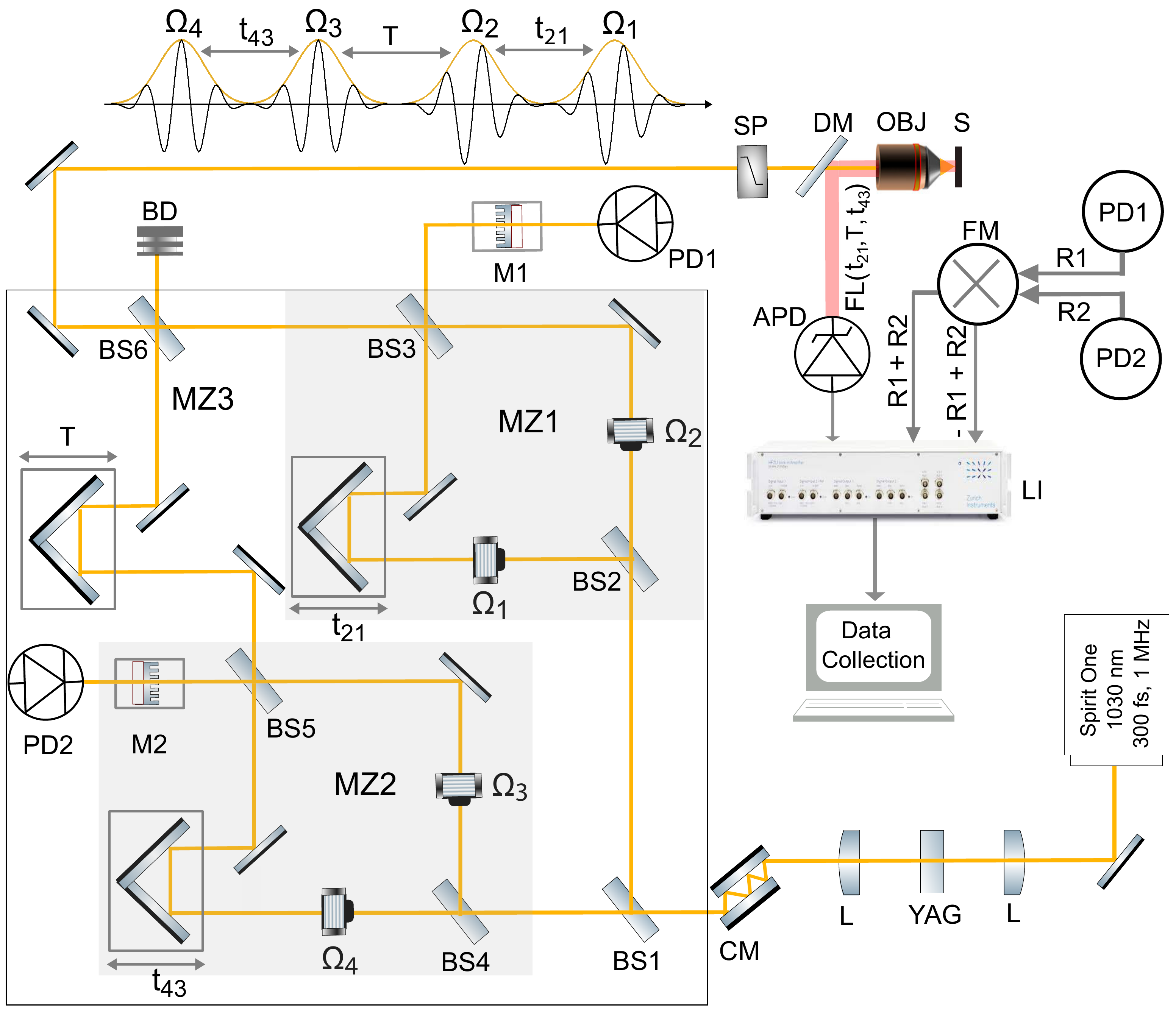}
	\caption{ Experimental schematic of the fMES spectrometer. Lens (L), chirped mirror (CM), beamsplitter (BS), acousto-optic modulator radio-frequency ($\Omega_s$), Mach-Zehnder interferometer (MZ), monochromator (M), photodiode (PD), beam dump (BD), shortpass optical filter (SP), longpass dichroic mirror (DM), reflective objective (OBJ), sample (S), avalanche photodiode (APD), frequency mixer (FM), electronic reference (R1, R2) and fluorescence (FL) signals, lock-in (LI) amplifier. The experimentally controllable positive time delays ($t_{21}$, $T$, $t_{43}$) between the pulses are shown in the figure.
	}
	\label{fig:fig1}
\end{figure*}

The schematic of the fMES experimental setup is shown in Fig.~\ref{fig:fig1}, and implements the AOPM approach first demonstrated by Marcus et al.\cite{Tekavec2007}. Approximately 1 $\mu$J, 300 fs pulses centered at 1040 nm from a 1 MHz Yb:KGW amplifier are focused into a 8 mm thick YAG crystal for generation of white light continuum (WLC). The WLC is then routed to two pairs of chirped mirrors (CMs, Layertec) for dispersion precompensation. A total of 34 pairs of bounces precompensates the expected optical dispersion in the setup.  The pulses are then split by a 50:50 broadband dielectric beamsplitter (BS, Newport), and each portion routed to two different balanced Mach-Zehnder (MZ) interferometers, MZ1 and MZ2. Within each MZ, the time-delayed replicas are tagged with a unique carrier-envelope phase which is cycled at radio frequencies using acousto-optic modulators (AOMs, Gooch and Housego). The phase modulation frequencies of the four arms are phase-synchronized by controlling them through a common clock (Novatech), and denoted as $\Omega_{i=1-4}$ in accordance with the pulse labels $1-4$. The time ordering of the pulses is also shown in the figure along with the corresponding time intervals $t_{21}$, $t_{32}$ or $T$, and $t_{43}$. When the split pulses within each interferometer, MZ1 and MZ2, are recombined at BS3 and BS5, respectively, the resulting pulse amplitudes modulate at the AOM difference frequencies, $\Omega_{12}$ and $\Omega_{34}$ for MZ1 and MZ2, respectively. $\Omega_{ij}$ is defined as $\Omega_{ij} = \Omega_i - \Omega_j$. Pulse pairs from each MZ are recombined at BS6 to form the third MZ interferometer, such that a collinear pulse train of 4 pulses, each with a unique carrier-envelope phase-shift, originates from BS6 for every laser shot. All time delays between the collinear pulses resulting after BS6 are controlled by translation stages (Newport MFA-CC for $t_{21,43}$, Newport ILS150CC for $T$ using Newport ESP301 stage controller).

As shown in the figure, one of the two output ports of each MZ is sent through a monochromator. The resulting output pulse of $\sim$1 nm width is centered at $\lambda_{R1,2} =$ 635 nm and corresponds to a full-width at half maximum (FWHM) time envelope of $\sim$0.6 ps. This ensures approximately constant reference amplitude as $t_{21}$ and $t_{43}$ time delays are scanned. Before the pulse pairs from MZ1 and MZ2 are recombined, MZ2 (or probe) pulses are both time-delayed by a waiting time $T$. The collinear train of 4 pulses is passed through a 675 nm shortpass filter (SP,Edmund Optics) and a 660 nm longpass dichroic filter (DM,Edmund Optics) before focusing into a sample using a (reverse Casegrain) reflective microscope objective (Thorlabs, LMM-40X-P01, NA=0.5). The fluorescence collection efficiency of this 0.5 NA microscope objective including the obscuration is $\sim$ 5 \%. The sample is circulated through a 100 $\mu$m pathlength flow cell (Starna) using a peristaltic pump (Masterflex) with sample flow rate of $\sim$85 ml/min. Sample stability over the entire data collection time was ensured by comparing optical density (OD) and shape of the absorption spectrum before and after the experiment (Section S4). The amplitude modulated fluorescence resulting from the sample, with frequency $\Omega_-$ and $\Omega_+$ for rephasing and non-rephasing signal pathways, respectively, is collected in the confocal geometry, transmitted through a 700 nm long pass filter (Edmund Optics), and detected using an avalanche photodiode (APD, Hamamatsu). The APD output  is sent into a lock-in amplifier (Zurich Instruments, HF2LI) for phase-sensitive detection relative to optically generated reference signals.\\

The optically generated reference signals $\omega_{R1}$ and $\omega_{R2}$ modulating at frequencies $\Omega_{12}$ and $\Omega_{34}$ are digitized and mixed using a 24-bit audio signal processor(Analog Devices) to generate mixed  reference signals modulating at frequencies $\Omega_{\mp}$. The mixed reference signals are connected as external references to the lock-in amplifier, and the rephasing and non-rephasing signals are detected in parallel channels using these references. As described in Section \ref{aopm}, such a detection scheme leads to physical undersampling of the optical frequencies at a shifted frequency $\omega_{eg} - \omega_R$, such that time intervals sampling optical coherences, $t_{21}$ and $t_{43}$, are less susceptible to mechanical fluctuations and optical coherences can be undersampled. In the experiments, the AOM frequencies are set at $\Omega_{1}$ = 80.109 MHz, $\Omega_{2}$ = 80.107 MHz, $\Omega_{3}$ = 80.005 MHz and $\Omega_{4}$ = 80.0 MHz through a common clock, such that non-rephasing and rephasing 2D signals oscillate at 7 kHz ($\Omega_{+}$) and 3 kHz ($\Omega_{-}$), respectively. Modulation of the resulting nonlinear signal at kHz frequencies partly minimizes the 1/$f$ noise. The powers per arm in each of the MZs are balanced by appropriately setting the AOM amplitudes so that on top of the microscope, we get an average power of 30 $\mu$W (corresponding pulse energy of 30 pJ) per interferometer.\\

\begin{figure*}[!ht]
	\centering
	\includegraphics[width=6 in]{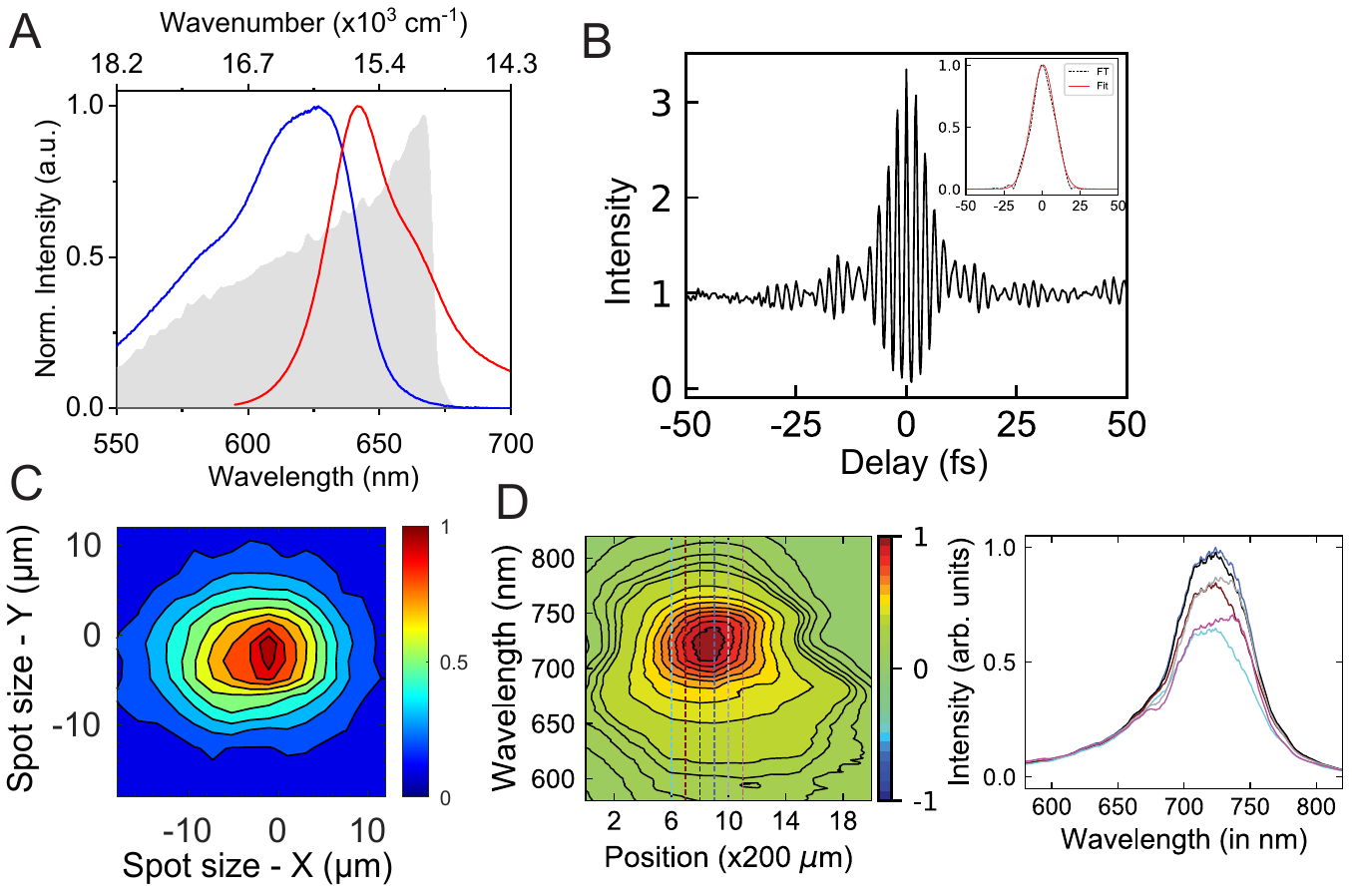}
	\caption{Pulse characterization. (A) Linear absorption (blue) and fluorescence (red) spectrum of Oxazine, overlaid with the measured laser spectrum (gray shaded area). (B) Experimentally measured collinear interferometric autocorrelation (IAC) trace (black) by placing a two-photon SiC photodiode at the objective focus. Fourier filtered trace of the IAC and fitted with an assumed Gaussian pulse (red, inset figure), which shows a FWHM pulse duration of $\sim$13 fs at the sample position. (C) Spot profile measured at the sample position using a monochrome CMOS sensor. Contours are drawn at the 5\% and 10-100\% in 10\% intervals. The raw image shown in Fig.~S1A. A 2D elliptical Gaussian fit yields FWHM spot sizes of $\sim$ 14 $\mu$m, 11 $\mu$m. The detailed description of the spot size measurement is given in Section S1. (D) Spatial chirp measurement. Normalized contour plot of the laser spectrum collected at each 200$\mu$m interval horizontal position of the spot transverse profile before it enters the shortpass filter. Contours are drawn at the 5$\%$, and 10-100$\%$ in 10$\%$ intervals for positive or negative contours. Dashed vertical lines are drawn within 60\% of the maximum amplitude in the spot transverse profile. Vertical slices at marked positions are plotted (right panel) to show the variation of the spectrum over the transverse profile.}
	\label{fig:fig2}
\end{figure*}

Widely employed pulse characterization methods such as frequency-resolved optical gating (FROG)\cite{Artigas2004, Steinmeyer2005}, multiphoton intrapulse interference phase scan (MIIPS)\cite{Dantus2008} rely on second-harmonic generation (SHG), where sufficient signal-to-noise (SNR) and SHG bandwidth are for key reliability of pulse compression. In the current case of WLC based fMES, this becomes challenging not only due to pJ pulse energies but also due to experimentally challenging collection of broadband UV spectrum through a microscope objective typically designed for the visible or near-infrared. On the other hand, a wide variety of materials which exhibit two-photon-induced photocurrent, have been shown to yield reliable results\cite{Wiersma1997, Riedle2000}, with maximum peak intensity attained for a transform-limited pulse. To measure the pulse duration, we used a pair of pulses from an interferometer and focus on a silicon carbide two-photon photodiode (SiC, Boston Electronics sg01XL-5ISO90) using the reflective microscope objective (OBJ) to record the two-photon autocorrelation signal. The diode was chosen so as to ensure that the spectral responsitivity curve overlaps well with the expected second harmonic of the laser spectrum. The resulting interferometric autocorrelation is shown in Fig.~\ref{fig:fig2}(B). By Fourier transforming the autocorrelation trace and numerically filtering out the oscillating components, we obtain a pulse duration of $\sim$13 fs at the sample position (Fig.~\ref{fig:fig2}(C)). Interferometric autocorrelation does not provide complete information about spectral phase, and we expect the minimum residual dispersion after compensation to be limited by one CM bounce or lesser (1 mm glass is compensated per bounce). To check this, we simulated the autocorrelation trace from the measured laser spectrum (gray shaded area of Fig.~\ref{fig:fig2}(A)) assuming a transform-limited pulse. From the simulated trace, the obtained pulse duration is $\sim$12 fs (Fig. S2). This is roughly consistent with the expectation that $\sim$21 fs$^2$ uncompensated dispersion, corresponding to less than one CM bounce, is sufficient to stretch the pulse duration from 12 fs to 13 fs. Note that the pronounced side peaks in the autocorrelation trace are observed both in the measured and simulated trace due to the sharp edge in the spectrum introduced by the shortpass and dichroic optical filters. We could not reproduce the expected 8:1 ratio in the autocorrelation which may be due to uncompensated higher-order dispersion in the setup. 

The spot size was characterized at the focus position after the reverse Casegrain reflective objective and shown in Fig.~\ref{fig:fig2}(D). The spot was measured by scanning a monochrome CMOS sensor (pixel resolution 2.08$\mu$m x 1.95$\mu$m) across the focus. The average FWHM focal spot size is $\sim$12 $\mu$m. Note that much smaller spot sizes could be obtained using glass objectives at the cost of excessive optical dispersion, or with reflective objective but at the cost of obscuring the core of the beam leading to insufficient pulse energies. Spatio-temporal chirp is expected in broadband setups replying on AOMs, such as the AOPM approached implemented here. Spatial filtering and AOM double-passing\cite{Jefferts2005} has been suggested\cite{Martin2020} to compensate for spatio-temporal chirp almost exactly, although at the cost of introducing optical dispersion and high power loss in the AOMs. Another way to minimize spatio-temporal chirp with lesser amount of extra optical dispersion and no extra power loss is through the use of prisms or lenses immediately after the AOM in order to prevent angular dispersion in the beam travel path. We have found the latter to be more suitable for the broadband WLC based setup with limited dispersion compensation ability presented here. We have checked this by characterizing the spatial chirp in the spot profile before routing it through the shortpass and dichroic filters, and the objective. This is shown in Fig.~\ref{fig:fig2}(E). The spot profile is $\sim$ 1.6 mm in diameter, and spectrum at each point of the profile is measured by scanning a fiber of core diameter200 $\mu$m (Thorlabs, M25L01) across this profile. The fiber coupler was placed on a mechanical translation stage (Thorlabs, PT1, least count 20$\mu$m), and the spectrum was taken at every  200$\mu$m interval. The spatio-temporal chirp at the focus is further minimized through achromatic focusing from a reflective objective. Minimization of spatial chirp implies that spectral phase imbalance in the MZ interferometers is expected to be minimal, such that, the imaginary component in the fluorescence excitation spectrum should be minimized. We confirm this further in Section \ref{phasing}.

\subsection{Data Collection Scheme}\label{collection}
\begin{figure*}[!ht]
	\centering
	\includegraphics[width=2 in]{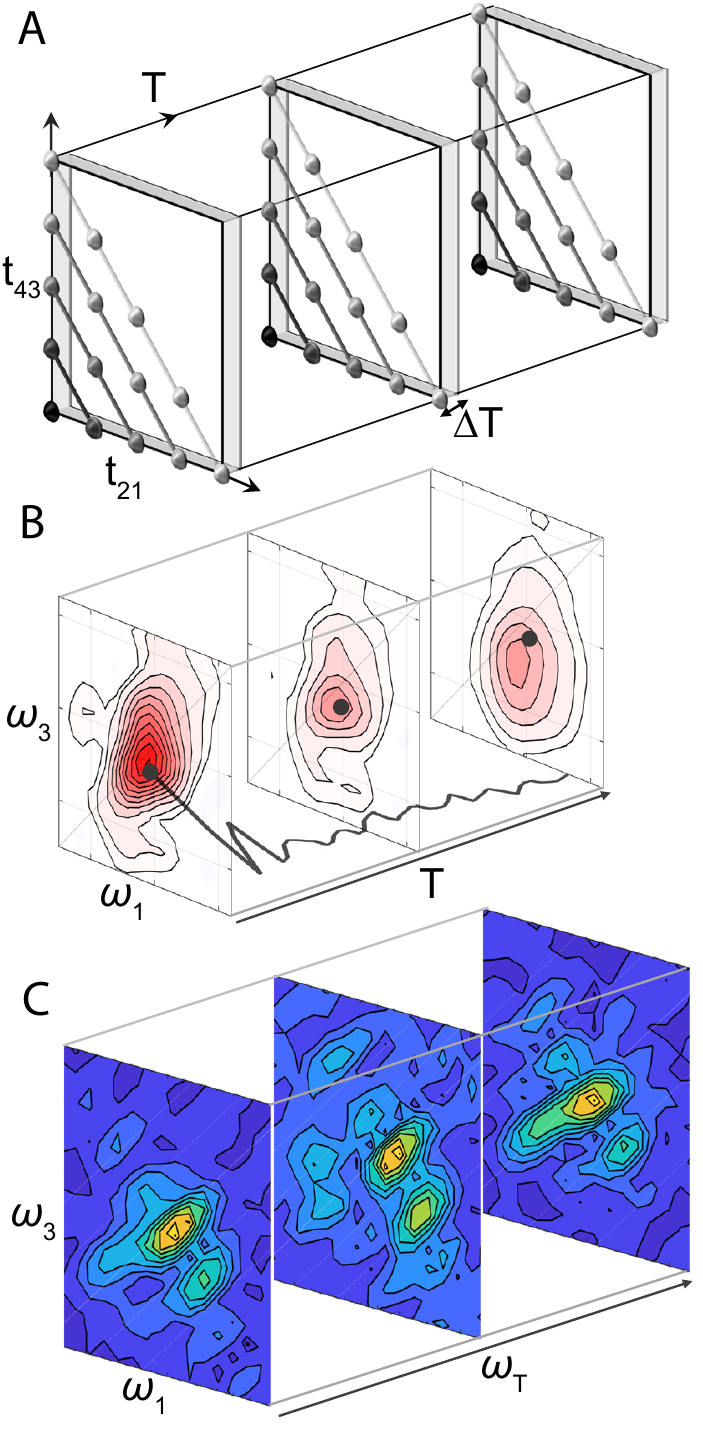}
	\caption{(A) Schematic for data collection and processing methodology. $t_{21}$ and $t_{43}$ points (shaded dots) are scanned step-wise, with points with similar shading scanned consecutively. For each ($t_{21,43}$) point, pump-probe delay $T$ is scanned continuously at a constant velocity. Multiple sweeps along $T$ scans are averaged first, followed by binning of averaged ($t_{21,43}$) data points into intervals $\Delta$T. (B) For each pump-probe delay $T$, Fourier transforms along $t_{43}$, $t_{21}$ yield a correlated map of detection versus excitation frequency. Each pixel of the 2D spectra in the resulting 3D dataset, depicted as gray oscillatory line, thus corresponds to a continuous $T$ scan. (C) The pixels are fitted globally with three exponential rate models to remove the exponential decay background, followed by Fourier transform of the residuals to obtain coherence maps along $\omega_{\text{T}}$ coherence axis.
	}
	\label{fig:fig3}
\end{figure*}

Construction of fMES dataset requires signal collection at each time delay along the delay axes ($t_{21}$,$T$, $t_{43}$). Even though noisy data points near zero delay affect the signal the most, typically delay scans for each $T$ have been done\cite{Tiwari2021} by sampling the optical coherence axes (corresponding to delays $t_{21}$ and $t_{43}$) as a uniform  ($t_{21}$,$t_{43}$) grid, through either stepwise or continuous stage scanning\cite{Ogilvie2021}, or through a pulse shaper\cite{Roeding2018}. Fig.~\ref{fig:fig3}(A) explains the alternative data collection scheme employed here. Stepwise sampling along the optical coherence axes ($t_{21}$,$t_{43}$) is biased towards the sampling delay points with equivalent time delays, rather than as uniform raster scanned grid. This is shown in Fig.~\ref{fig:fig3}(A), where data points of darker shade are scanned first and data points of the same shade are scanned consecutively. With this scheme, the data points at which the signal S($t_{21}$,$t_{43}$) is expected to be the weakest are scanned at the very end, thus making the measurement less susceptible to long-term laser drifts. Following the above scheme, the time delays are scanned stepwise from 0 to 51 fs in step of 3 fs, corresponds to a Nyquist frequency of $\sim$5556 cm$^{-1}$. The signal level $t_{12,34}$ = 51 fs is already below 5\% of the maximum signal, implying a system limited frequency resolution in the Fourier transformed 2D spectra. Another crucial departure from the existing sampling schemes is that the pump-probe delay ($T$) is scanned continuously from -50 fs to 1 ps with a uniform stage velocity of 20$\mu m$/sec, corresponding to 7.5 secs/sweep/$t_{21,43}$ data point. This minimizes the sample exposure window substantially compared to existing approaches, to provide a scalable route towards spatially-resolved measurements where reducing sample exposure window is highly desirable. Signals are digitized at a sampling rate of 7.2k samples/sec with a lock-in time constant of 5 ms. With an acceleration of 20 mm/sec$^2$, the stage reaches the set constant velocity in a distance of the order of minimum incremental resolution of the stage. Step scans along $T$ for a fixed $t_{21,43}$ delay were overlaid with continuous $T$ scans in order to ensure the reproducibility between the two (Fig.~S5). Scanning from -50 fs allows us to accurately determine zero $T$ delay by averaging multiple $t_{21,43}$ sweeps. This is described in Section \ref{phasing}.

In the above scheme, minimum experimental time resolution along $T$ will be determined by the stage velocity and the lock-in time constant. Allowing the lock-in signal to settle to >99\% of the actual value (5x time constant), the time resolution along $T$ is 3.3 fs. Total 4 sweeps along increasing $T$ direction are averaged. A slower lock-in time constant or a faster $T$ velocity makes $T$ resolution poorer, although the latter also makes the experiment faster and could be done to further optimize the experiment and minimize the light exposure window.

Collection of one full 3D data set takes $\sim$40 minutes with one $T$ sweep per ($t_{21}, t_{43}$). For the 2DES data reported in Section \ref{results}, total 4 $T$ sweeps per ($t_{21}, t_{43}$) are performed. In comparison, stepwise scans along $T$ at 1 MHz repetition rate, with substantially poorer SNR compared to 83 MHz\cite{Tiwari2018a}, would have taken $\sim$4.5 hours for one $t$ scan, assuming the same time resolution, scan range and stage wait time of 150 ms (lock-in time constant 30 ms). Due to long data collection times with $T$ axis being the slowest, step scanning $T$ makes measurement of weak coherences along $T$ most susceptible to laser drifts. In case of continuous $T$ scanning, reduced data collection time and therefore less susceptibility to laser drifts, multiple sweeps along $T$, and the ability to bin multiple $T$ samples to get a well averaged $t_{21,43}$ data point (Section \ref{processing}), are expected to contribute to higher SNR. Fine sampling along $T$ has also been suggested\cite{Dahlberg2013, Dahlberg2017,Gioro2018} to effectively average out high-frequency noise in the data.

\subsection{Data Processing}\label{processing}
As shown in Fig.~\ref{fig:fig2}A, multiple sweeps along $T$ are averaged first, followed by binning of averaged $T$ sweep into intervals $\Delta T$ = 10 fs. For 10 fs $T$ steps, a total of $\sim$540 samples fall within each $T$ bin as determined by the sampling rate and stage velocity. These data points are averaged together to form one $t_{21,43}$ averaged data point. Overall, the averaging and binning procedure yields uniformly spaced $T$ delay points, with a 3D grid of 18$\times$18$\times$101 time points along $t_{21}$, $t_{43}$ and $T$ delay axes, respectively.

In the AOPM approach implemented here, simultaneous collection of both rephasing and non-rephasing signals, modulating at unique modulation frequencies, $\Omega_{-}$ and $\Omega_{+}$, respectively, is possible through two parallel lock-in channels referenced to corresponding optically generated and digitally mixed frequencies. At each pump-probe waiting time $T$, the data is zero-padded up to 64 grid points, prior to Fourier transform along $t_{21}$ and $t_{43}$. Note that zero-padding up to 2$\times$18 number of points is allowed\cite{Ernst1973} by Kramer-Kronig relations and yields frequency resolution of $\sim$ 327 cm$^{-1}$ along the corresponding $\omega_{1,3}$ axes. Increasing the number of points from 36 to 64 (nearest power of 2) interpolates the data without enhancing  spectral resolution which is system limited due to fast optical dephasing within $\sim$50 fs. 2D Fourier transform of the complex dataset yields a correlated map of  detection frequency (Fourier transform of $t_{43}$ to $\omega_3$) versus excitation frequency (Fourier transform of $t_{21}$ to $\omega_1$), for a given $T$. Rephasing and non-rephasing spectra are phased separately and added in the frequency domain to yield absorptive 2D spectrum. The phasing procedure is discussed in Section \ref{phasing}. 

As depicted in Fig.~\ref{fig:fig2}B, a given pixel in the 2D plot may have contributions from coherent and incoherent signal pathways along $T$. For isolating the coherences in the collected data, the 3D rephasing dataset is fit globally to a tri-exponential fit which removes incoherent exponentially decaying population signals from the data. The 3D dataset can be fit from $T$ = 0 fs because elimination of non-resonant signals and scatter in fluorescence-detection makes early $T$ signals easily accessible in fMES. The resulting residuals are zero padded from 101 to 256 points along $T$ prior to Fourier transform, although the expected frequency resolution of $\sim$16.5 cm$^{-1}$ along $\omega_T$ corresponds to only twice the range scanned along $T$.  This results in a three-dimensional frequency cube $S(\omega_{1},\omega_{T},\omega_3)$, shown in Fig.~\ref{fig:fig3}(C) with coherence frequency $\pm\omega_T$ corresponding to $T$. The total $\omega_T$ frequency content in the data is given by the Frobenius norm defined as ${F(\omega_T)} = \abs{\sqrt{\sum_{i,j}\abs{S(\omega_{1,i},\omega_{T},\omega_{3,j})}^2}}$. $F(\omega_T)$ reports on the most prominent frequencies in the dataset as well as the noise floor of the collected data. For a given coherence frequency $\omega_T$, plotting the absolute value of the Fourier transformed signal as $\abs{S(\omega_{\tau},\omega_t;\omega_T)}$ is referred to as a coherence map (CM) at $\omega_T$, and reports 2D positions of the most prominent quantum beating amplitudes. With sufficient experimental SNR and phase stability, further separation of coherent signal pathways contributing to CMs is possible by starting from the complex rephasing dataset. The resulting sign of the coherence frequency, $\pm\omega_T$, is physically significant and can be used to isolate various ground and excited state quantum beat pathways\cite{JonasARPC2018}. This is demonstrated in Section \ref{results}.

\subsection{Determination of Zero Delays and Phasing}\label{phasing}
The zero time delays for $t_{21}$ and $t_{43}$ optical coherence axes are determined by collecting the linear fluorescence signal separately for each interferometer MZ1 and MZ2, where the signals  are  modulated at the difference AOM frequencies $\Omega_{12}$ and $\Omega_{34}$ respectively. The respective zero time delay for an interferometer is determined by the maximum absolute value of the demodulated linear signal by scanning the respective delay from -60 fs to +60 fs in 1 fs steps, with stage settling time of 200 ms (greater than 5 times the lock-in time constant). Multiple trials are repeated to determine an average zero time delay position along each delay axis with an error bar of $\sim$0.2 fs, which is within the minimum incremental delay resolution (0.7 fs) possible with the stage. The $T$ zero delay is obtained by scanning the fixed arm of each interferometer (arms 2 and 3) with modulation frequency $\Omega_{23}$. Other combinations, for example between arms 1-4 and 1-3 and 2-4, were also verified to yield the same zero $T$ delay. Since $T$ delay is scanned continuously during 3D dataset collection, the zero delay position obtained above is only a reference time point for phasing the lock-in signal (see following paragraph), while $T$ zero delay point needs to be re-determined for the continuous scan. As discussed in Section \ref{collection}, continuous scans from -50 fs for each grid point ($t_{21}$, $t_{43}$) can be overlaid to determine an average $T$ = 0 fs position. However, fast optical dephasing along $t_{21,43}$ implies that this can be reliably done only for the first few time delay points (darker shades in Fig.~\ref{fig:fig2}(A). This is shown in Fig.~S4 in SI.

Spectral phase imbalance in the interferometer, for example due to spatial chirp can lead to residual phase $\mp \Delta\phi_{21}(\omega_1) + \Delta\phi_{34}(\omega_3)$ in the rephasing and non-rephasing signals (Section \ref{aopm}). The residual phase mixes the real and imaginary lineshapes to yield mixed absorptive and dispersive lineshapes in the absorptive 2D spectra.  In case of time-domain fMES implementation discussed here, interferometer drifts or zero-delay errors in MZ1 and MZ2, denoted as $\delta t_{21,34}$, cause phase drifts $\mp\Delta\omega_{1}\delta t_{21} + \Delta\omega_2\delta t_{34}$, in rephasing and non-rephasing signals, respectively. Here $\Delta\omega_{1,2}$ is defined as $\omega_{eg} - \omega_{R1,2}$. The resulting phase drifts also mix real and imaginary lineshapes in the individual channels.

In conventional MES, the total residual spectral phase is approximated as  $\Delta\phi(\omega) = \phi_o + \omega\delta t + \phi''\omega^2/2$, and can be corrected in the frequency domain along the detection axis, assuming no residual phase along the excitation axis. In the AOPM approach, contributions from both, $\Delta\phi_{21}(\omega_1)$ and $\Delta\phi_{34}(\omega_3)$, are expected. Arbitrary constant phase due to lock-in detection is set to zero at the start of the experiment, while $\omega_{R1,2}$ can be chosen to minimize phase drift contributions. In order to correct fMES spectra for interferometer imbalance arising from spatial chirp, Agathangelou et al. have derived\cite{Ogilvie2021} the residual spectral phase, $\Delta\phi_{21}(\omega)$ and $\Delta\phi_{34}(\omega)$, by collecting the linear fluorescence signals from MZ1 and MZ2 (Eqn.~\ref{eq2} of Section \ref{aopm}), respectively. Symmetrically scanned linear fluorescence signal is expected to be completely real, hence a substantial relative magnitude of the imaginary part would indicate spectral imbalance within the interferometer which can then be corrected. In our experience, constant phase factors $\phi_o^{R, NR}$, to individually mix real and imaginary parts in the rephasing and non-rephasing channels, are already sufficient to phase the absorptive 2D spectra fairly well for all $T$. This is shown in Fig. S6 where the residual spectral phase derived from the linear signal is unable to correctly phase the 2D spectrum. This observation is consistent with the negligible spatial chirp across the beam spatial profile (Fig.~\ref{fig:fig2}E) and $\sim$10x weaker imaginary component in the Fourier transform of linear fluorescence signals (Fig. S6), which together indicate minimal spectral imbalance in the interferometers.

\section{Results and Discussion}\label{results}
In this section, we discuss the sensitivity of fluorescence-detection followed by an experimental demonstration which serves as a measure of sensitivity. Following the approach described in Section \ref{collection}, high SNR detection of 2D spectra down to 2-3 orders of magnitude lower OD than typical approaches is presented. This is extended one step further to demonstrate phase-sensitive detection of vibrational quantum beats in the form of 2D CMs of a laser dye Oxazine 720 in ethanol at room temperature. The sample preparation at different concentrations, linear absorption and fluorescence spectra of Oxazine 720 in ethanol are described in Section S3. 

\subsection{Sensitivity of fMES}
In the context of pump-probe measurements, Cho et al. have derived and experimentally verified\cite{ChoAbsoluteSignal,ChoAllOptical} expressions for absolute photon number changes in the probe beam caused by the pump. These expressions are consistent with the expression\cite{Hybl2001} for relaxed 2D spectrum incorporating propagation distortions caused by attenuation of excitation pulses and the signal as they travel through a sample, and serve as a useful starting point to think about the dependence of fMES signal on the experimental parameters. The non-linear fluorescence signal $S_f$ for a two-electronic level system can be simply expressed as --
\begin{equation}
S_f \sim f_{rep}\Phi [\rho_u (x,y,z) \delta . N_o\sigma_{ge}L] [\rho_r(x,y,z) \delta (\sigma_{ge} + \sigma_{eg})]\eta_{det}.
\label{eq4}
\end{equation}
In the above equation, $f_{rep}$ is the repetition rate, $\Phi$ is the quantum yield which is assumed to be excitation frequency-independent, $N_o$ is the molecular number density in the ground electronic state, and $L$ is the sample pathlength. Polarization and $T$ lifetime dependence has been ignored. The simplified Eqn.~\ref{eq4} assumes approximately transparent samples with no fluorescence re-absorption, identical pump pulses, as well as identical probe pulses. Pump and probe pulses are denoted by subscript `u' and `r' respectively. The pump and probe spectrum is assumed to be nearly monochromatic with width $\delta$. As in ref.\cite{ChoAllOptical}, the first term in the square brackets arises from total number of molecules excited by the pump as it travels through the sample. The term in the second square bracket arises from the overlap of probe pulse spectrum with the absorption and emission cross-sections. Although pump and probe are always fully overlapped in a collinear geometry, an additional assumption in Eqn.~\ref{eq4} is that of constant pump and probe spotsizes within the sample pathlength, such that the entire pathlength $L$ contributes to the signal. This assumption does not hold when focusing using a microscope objective and only a smaller region within $L$ where the fluence is maximum dominantly contributes to the nonlinear signal. For example, ref. \cite{Tiwari2018a} has discussed that the majority of the non-linear signal along the axial direction is generated from within the point spread function (PSF) created by the objective. This can also be incorporated explicitly in the fluence $\rho(x,y,z)$ by including the transverse profile dependence as the beams contract, focus and expand within the 100 $\mu$m sample pathlength. However, we have checked the validity of the above assumption. This is shown in Fig.~S1. The last factor $\eta_{det}$ in Eqn.~\ref{eq4} represents the overall loss of time-integrated fluorescence signal in the detection line, which includes overlap of longpass optical filter with the steady-state fluorescence spectrum of the sample, fluorescence collection efficiency of the objective, and spectral responsitivity and gain of the APD. White noise and pink ($1/f$) noise can be minimized through lock-in frequency filter and averaging, and phase-modulation at high frequencies, respectively. The noise in data collection at a given modulation frequency also depends on how the APD gain-dependent noise spectral density\cite{apdnoise} compares with the signal size $S_f$. All such noise contributions together determine the signal-to-noise ratio (SNR) of the experiment. Overall, Eqn.~\ref{eq4} indicates that $S_f \approx \sigma  OD$ and that a fair comparison of experimental sensitivity must ensure that $f_{rep}$, $\Phi$, $\eta_{det}$, fluence $\rho$ and absorption cross-section $\sigma$ are comparable. Note that the fluence per pulse used in our experiment is $\sim$12$\mu$J/cm$^2$, corresponding to a pump excitation probability\cite{ChoAbsoluteSignal} of < 1\% at the center of the beam profile.

With above considerations, signal size is expected to be proportional to OD, for ODs $\ll$ 1. Conventional 2DES measurements have been conducted at an OD of $\sim$0.3 where the four wavemixing signal maximizes\cite{Hybl2001,ChoAbsoluteSignal}. Hence as a first demonstration of the continuous $T$ scanning approach to fMES, we present the absorptive 2DES spectra at ODs $\sim$2-3 orders of magnitude lower than what is typical.

\subsection{2D Spectra Versus Sample OD}\label{2D}
Oxazine 720 (also known as Oxazine 170) is an interesting laser dye in the Oxazine family commonly used as near-IR fluorescent probes for bio-imaging applications. Oxazines are derived from Xanthenes by replacing the central carbon with a nitrogen atom which, unlike the central carbon, participates in $\pi$-conjugation resulting in $\sim$100 nm red-shift of the $S_0 \rightarrow S_1$ transition. Oxazine 720 has fluorescence quantum yield\cite{Reisfeld1989} of $\sim$0.53 in ethanol with absorption and fluorescence peaks located at 627 nm and 641 nm, respectively(Fig.\ref{fig:fig2}(A)). The blue shoulder in the absorption spectrum is expected to be a Franck-Condon (FC) progression\cite{Kostjukov2021}. N-H vibrations, rotations of the aminoalkyl substituents, or radiationless energy transfer from the solute to high-frequency combination bands of hydrogen-bonded solvents are typically invoked\cite{Drexhage1976,Maillard2021} to explain radiationless deactivation of the excited electronic states in Oxazine dyes. In case of Oxazine 720, a fused benzene substituent on the skeleton partially limits the rotation of aminoalkyl substituent to inhibit\cite{Drexhage1981} radiationless deactivation. Oxazine 720 is also known to exhibit solvatochromism, with Stokes' shift increasing with solvent polarity. In polar solvents such as methanol the ion-pair in Oxazine 720 breaks and the cationic skeleton is expected to be less rigid, although Condon approximation seems to hold\cite{Baumgartel1988}. It is reported\cite{Zaker2015} that upon photoexcitation, the permanent dipole moment of Oxazine 720 increases substantially in the excited state. Ref.\cite{Kostjukov2021} has also reported increased strength of solute-solvent H-bonding upon electronic excitation. Thus, rich photoinduced solvation dynamics is expected for this laser dye and will be a subject of future investigations in our group.

\begin{figure*}[!ht]
	\centering
	\includegraphics[width=5 in]{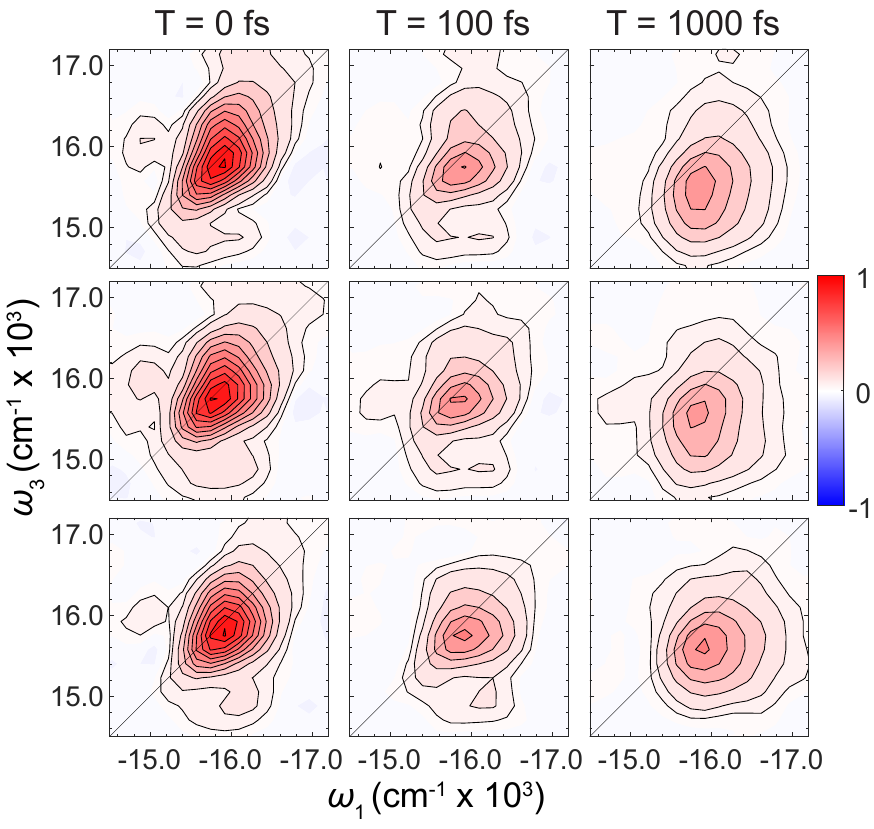}
	\caption{ Real absorptive two-dimensional electronic spectra of Oxazine 720 in ethanol in 100 $\mu$m pathlength flow cell, at the pump-probe waiting times $T$ = 0 fs, $T$ = 100 fs, and $T$ = 1 ps. The sample OD is (A) 96 mOD, (B) 9.5 mOD and (C) 1.4 mOD. Contours are drawn at the 5$\%$, and 10-100$\%$ in 10$\%$  intervals for both positive or negative contours.
	}
	\label{fig:fig4}
\end{figure*}

Fig.\ref{fig:fig4} shows the experimentally measured absorptive fMES 2D spectra for representative pump-probe delays $T$ = 0 fs, 100 fs, 1000fs. The three rows in the figure compare the room temperature 2D spectra at three different sample ODs ranging from $\sim$ 1 mOD to 100 mOD. In terms of experimental parameters, the three datasets only differ in terms of sample OD and the APD gain, ignoring the day to day variations in experimental conditions. Rich solvation dynamics is evident in the 2D spectra with rapid loss of frequency correlation and red-shifted detection frequency (Stokes' shift) seen within $\sim$1 ps. The 2D spectra and features associated with solvation dynamics are consistent to within $\sim$10\% contour across the range of ODs measured, indicating high-sensitivity of detection at ODs $\sim$2-3 orders of magnitude lower than what is typical in 2DES. 

Starting with samples of $\sim$300 mOD optimized\cite{ChoAbsoluteSignal,Hybl2001} for a maximum signal size, pump-probe measurements have routinely reported signal amplitudes of $\sim$10 mOD. Optimized pump-probe spectrometer with a parallel reference spectrograph and shot-to-shot detection has reported\cite{Ernsting2010,Lang2018} noise baselines of the order of 1 mOD (10$^{-3}$). Compared to these earlier works, measurements of consistent 2D spectra and solvation dynamics reported here are significant because the \textit{starting} sample OD is as low as $\sim$ 1 mOD. Recently Tokmakoff and co-workers have also reported 2DIR spectra at concentrations $\sim$1/2 of typical, where the sensitivity enhancement is achieved by encoding mid-IR transitions as electronic excitations and detection the resulting fluorescence\cite{tokma2018}.   

\subsection{Coherence Maps Versus Sample OD}
Amplitude modulations resulting from intramolecular vibrational wavepackets at room temperature are typically only a few percent of the incoherent population. For example, a FC displacement of half the zero-point amplitude results in 2D oscillation amplitude\cite{Jonas2008vibr} of $\sim$12.5\% of the maximum 2D signal expected from a monomer. 2D CMs can isolate\cite{Butkus2017} excited state wavepackets based on quantum beat phase along $T$. However, this requires high SNR detection of vibrational wavepackets at room temperature. This becomes especially challenging with a WLC based setup where 5\% RMS spectral fluctuations across the entire spectral bandwidth are typical\cite{Ernsting2010,Bradler2009,Bradler2014} compared to light sources based on multi-stage optical parametric amplifiers(OPAs) operating in the saturation regime\cite{Kartner2009}. Thus, a further demonstration of the sensitivity of continuous $T$ scanning approach to fMES would be a measurement of CMs from weak vibrational wavepackets along with the quantum beat phase at the lowest sample ODs.

Conventional 2DES experiments\cite{Turner2019,Weng2020} on the Oxazine family of dyes have reported a prominent intramolecular Raman-active mode around $\sim$586 cm$^{-1}$, associated with in-plane bending motion of the phenyl ring. Turner et al. have also implicated\cite{Turner2016} this mode as the coupling mode responsible for coherent surface crossing mediated by a conical intersection in the related laser dye Cresyl Violet.  Their quantum-chemical computations\cite{Turner2022} suggest that the prominent vibrational mode of $\sim$586 cm$^{-1}$ exhibits much larger Huang-Rhys (HR) factor of $\sim$ 0.28 than other modes present in the system. The 2DES vibrational wavepacket modulation from this HR factor is expected to be only $\sim$1/3 of the incoherent population background. In comparison, for acene family of molecules such as pentacene, HR factors > 1 are typical with facile detection of dominant vibrational wavepackets at room temperature\cite{Tan2021}.

\begin{figure*}[!ht]
	\centering
	\includegraphics[width=3 in]{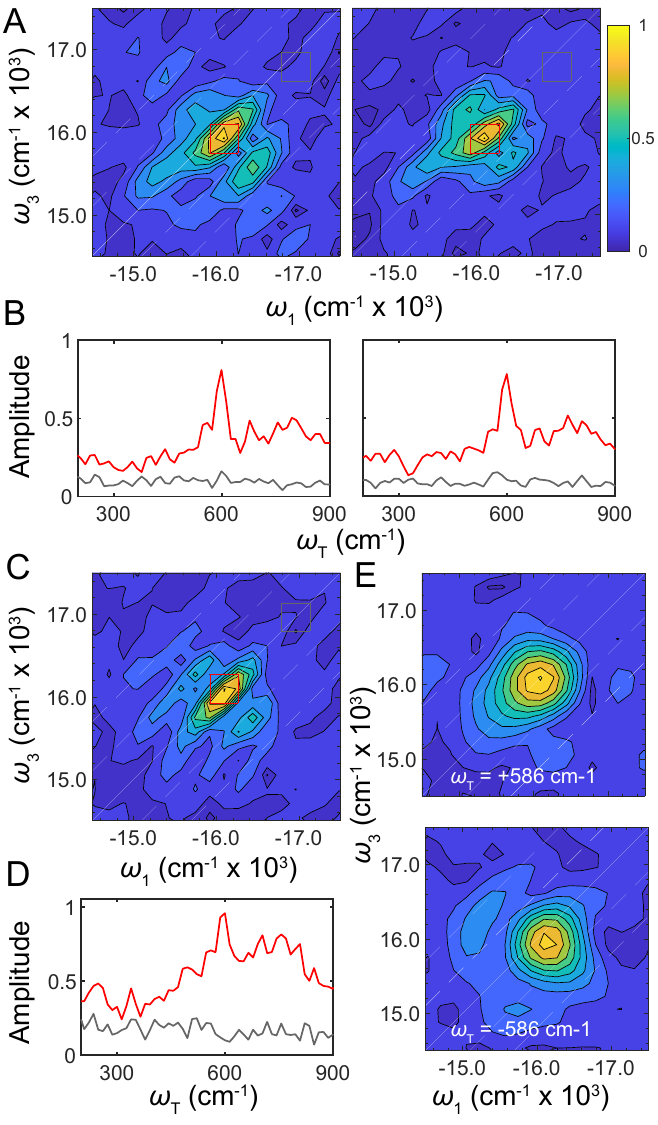}
	\caption{(A) Real rephasing coherence maps (CMs) for $ \omega_T = 586$ cm$^{-1}$ measured with a sample OD of 96 mOD (left) , 9.5 mOD (right). The diagonal line in the CMs correspond to the 2D diagonal and the dashed lines are separated by one quantum of $\omega_T$ vibration (586 cm$^{-1}$). Contours are drawn at the 5$\%$, and 10-100$\%$ in 10$\%$ intervals. The CMs are normalized to their corresponding maxima. Red squares are drawn around the maximum amplitude of the map, and the gray squares are drawn at a position of almost two vibrational quanta away from the red squares. (B) Frobenius spectra are calculated over the above red (signal) and gray (noise) areas. (C) Real rephasing CM for $ \omega_T = 586$ cm$^{-1}$ measured with a sample OD of 1.4 mOD and the corresponding Frobenius spectra is shown in panel (D). (E) Complex rephasing CM $+\omega_T$ (top), $-\omega_T$ (bottom) for the vibrational frequency of 586 cm$^{-1}$ measured at a sample OD of 1.4 mOD. $\pm\omega_T$ maps are normalized with respect to the common maxima of two maps.} 
	\label{fig:fig5}
\end{figure*}

Fig.~\ref{fig:fig5} compares real rephasing CMs and associated SNR for measurements on Oxazine 720 with sample ODs ranging from $\sim$1-100 mOD. As mentioned in Section \ref{2D}, only APD gain is changed between the experiments while the number of $T$ averages are the same between the scans. The experimental CMs of the most prominent vibrational mode at $\omega_T =$ 586 cm$^{-1}$ are shown in Fig.~\ref{fig:fig5}(A). Maximum CM amplitude for a given OD also corresponds to the maximum amplitude coordinate in the respective 2DES spectrum at $T = 0$ fs. We also observe that the maximum CM amplitude is $\sim$ 1/4 of the maximum 2D amplitude (see Table S1). This suggests that the FC displacement of the 586 cm$^{-1}$ mode may be smaller for Oxazine 720 in ethanol than that for Cresyl Violet as reported by Turner et. al.\cite{Turner2022}. The CM amplitude positions are approximately consistent with a displaced harmonic oscillator model of one vibrational frequency mode at 586 cm$^{-1}$. 

As shown in Fig.\ref{fig:fig5}(C), the CM features for the lowest sample OD ($\sim$1 mOD) are fairly consistent (to within $\sim$ 20\%) with that of the highest OD. Fig.~\ref{fig:fig5}(B),(D) show the SNR analysis for each of the CMs collected across different sample ODs by evaluating the Frobenius norm over the main CM amplitude region and a noisy region where no CM amplitude is expected. Both regions are marked as squares in the figure and consistent across the three cases. It is seen that the SNR of $\sim$8.5 for $\sim$100 mOD and $\sim$10 mOD cases does not deteriorate, whereas, for the case of 1 mOD, the SNR deteriorates to $\sim$5.5. The noise floor deterioration at the lowest sample OD may be expected due to a larger APD gain\cite{apdnoise} degrading the spectral noise floor. 

 In a real rephasing coherence map vibrational wavepackets from both excited state and ground states contribute at the diagonal peak such that the diagonal CM amplitude represents the sum of both pathways which can be separated based on the quantum beat phase along $\omega_T$\cite{Tiwari2018b}. Fig.\ref{fig:fig5}(E) isolates excited state contributions by plotting rephasing 2D CMs at $\pm \omega_T = 586$ cm$^{-1}$ for the lowest sample OD. The location of the peak positions can be explained\cite{Tiwari2018b} through Feynman pathways considering a displaced harmonic oscillator model, such that $+\omega_T$ contributions are expected to arise only from the excited state vibrational quantum beats. Although the individual CM peak contributions are merged together, diagonal contributions seem more prominent compared to other 2D locations. The resulting $\pm \omega_T$ CMs suggest that an SNR of $\sim$5.5 is good enough for isolating excited state ($+\omega_T$) contributions with a respectable SNR. Overall, consistent CM measurements at ODs 2-3 orders of magnitude lesser than typical demonstrates the sensitivity limits of the approach implemented here. It should also be pointed out that even though the measurement sensitivity reported here is high enough to yield an SNR of 5.5 for the most prominent mode at ODs 2-3 orders of magnitude lower than typical, only one prominent mode is consistently seen in the experiments suggesting overall sensitivity limitations in constructing the $\omega_T$ dimensions despite $T$ averaging.

\begin{figure*}[!ht]
	\centering
	\includegraphics[width=4 in]{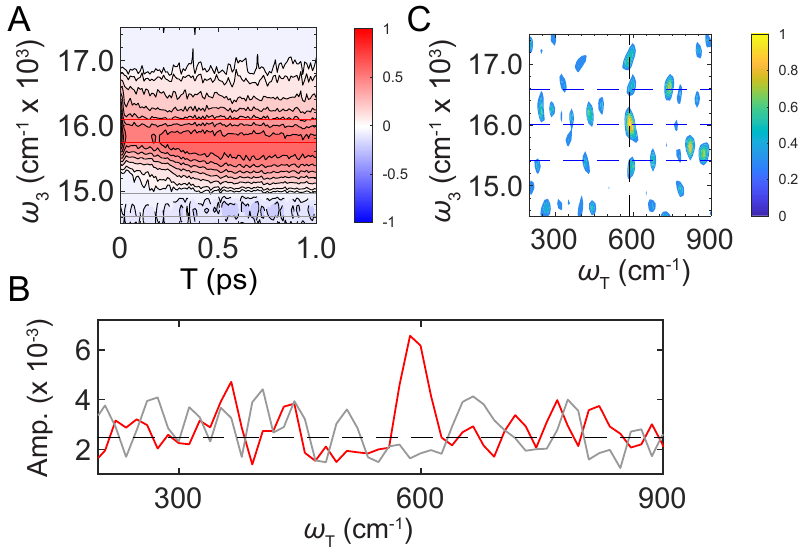}
	\caption{(A)Normalized real rephasing spectrally resolved pump-probe (SRPP) data for 1.4 mOD sample of Oxazine 720 at room temperature averaged over 8 $T$ sweeps. Contour lines are drawn at the 5$\%$, and 10-100$\%$ in 10$\%$ intervals for both positive or negative contours. (B) Frobenius spectra were calculated over red and gray regions marked in panel A. The prominent vibrational frequency at $\omega_T$=586 cm$^{-1}$ was also seen in the 2D CMs. A horizontal dashed line is drawn at $\sim$ 30\% of the maxima value of the Frobenius spectrum to denote the average noise floor. (C) Normalized absolute squared Fourier transform map of the SRPP data obtained after a three-exponential global fit of the population kinetics. Here the contours are drawn from 30\% to 90\% in steps of 10\% and 90\% to 100\% in 2\% intervals. A vertical dashed line marked at $\omega_T$=586 cm$^{-1}$ shows the prominent vibrational frequency. The central dashed horizontal line is drawn at a position corresponding to the 2D CM maximum, and the other two lines are drawn at one vibrational quantum above and below the main peak.}
 	
	\label{fig:fig6}
\end{figure*}

More number of $T$ sweeps is expected to improve the SNR for the case of $\sim$1 mOD although at the cost of increased experimental time. Instead, a significantly faster experiment, similar to a spectrally resolved pump-probe (SRPP) experiment\cite{JonasARPC2003}, can be performed by fixing $t_{21} = 0$ fs. The rest of the parameters are exactly the same as in Section \ref{collection}. A related fluorescence-detected pump-probe (fPP) experiment was recently reported by Maly et al.\cite{Maly2021}. Although excitation frequency information is not available in a pump-probe experiment, high-sensitivity detection of coherent wavepackets at the lowest ODs, collection of both rephasing and non-rephasing channels and isolation of pathways based on quantum beat phase is still possible in the AOPM approach to fPP taking it beyond conventional pump-probe spectroscopy.  Pump-probe delay $T$ was scanned continuously from 0 to 1 ps, with multiple sweeps for each $t_{43}$ delay. $t_{43}$ delay was scanned from of 0-51 fs in steps of 3 fs. One ($T,t_{43}$) scan now takes only $\sim$2.25 minutes compared to 40 minutes in the 3D scan. The resulting 2D data set is then Fourier transformed along $t_{43}$ axis to yield the detection axis ($\omega_3$). The final SRPP spectrum for the rephasing pathways is shown in Fig.\ref{fig:fig6}(A) where a total of 8 $T$ sweeps per $t_{43}$ time point were performed for the reported SNR. Polar solvation dynamics evident from changing 2D lineshapes (Fig.~\ref{fig:fig4}) is seen as ultrafast red-shift along the detection axis in Fig.~\ref{fig:fig6}(A). Similar to 3D scan, a 3-exponential global fit of the data, and Fourier transform of residuals yields a 2D map of $\omega_3$ and $\omega_T$ , shown in Fig.\ref{fig:fig6}(C). The corresponding Frobenius map (Fig.~\ref{fig:fig6}(B)) calculated over the frequencies shown in red lines in Fig.~\ref{fig:fig6}(A), shows the prominent coherence frequency $\omega_T =$ 586 cm$^{-1}$ also seen in the 2D CMs (Fig.~\ref{fig:fig5}). Similar to Fig.~\ref{fig:fig5}(B), the Frobenius norm is compared with the noise floor calculated over frequencies marked by gray lines in panel A, resulting in SNR comparable to the 2D CMs but with a significantly faster experiment. The corresponding $(\omega_T,\omega_3)$ map in panel C shows the location of the 586 cm$^{-1}$ mode (vertical dashed line). The location of the main 2D CM peak amplitude (Fig.~\ref{fig:fig5}C), shown as dashed horizontal line, is consistent with the corresponding peak location in the SRPP spectrum. Faint amplitudes are seen one vibrational quanta above and below the main peak, although with magnitude comparable to the noise floor. The SRPP spectrum demonstrates SNR comparable to 2D CMs along with the ability to isolate signals based on $T$ quantum beat phase, but with a significantly improved experimental time. The flexibility of continuous $T$ scanning approach also allows for more averages, faster scan velocity, and reconstructing the $\omega_1$ axis as well, suggesting that a slower 2D CM experiment may not be that advantageous overall.

\subsection{Compressive Sensing Enables Faster Experiment}
\begin{figure*}[!ht]
	\centering
	\includegraphics[width=6 in]{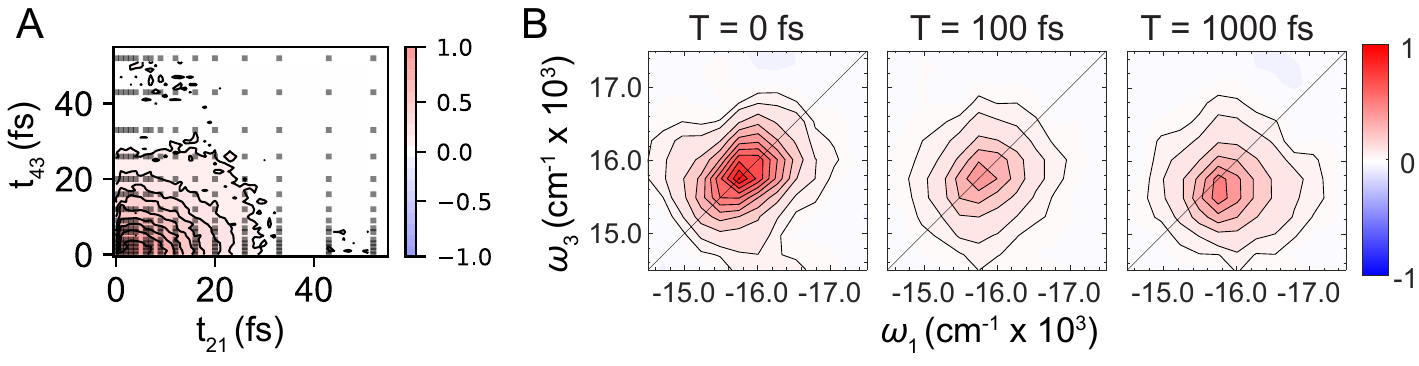}
	\caption{(A) Exponential sampling (ES) scheme applied on a 52 $\times$ 52 dense uniform grid  leads to only $15 \times 15$ number of points with maximum sampling frequency in the regions where the signal is maximum. Experimentally measured time domain signal for the dense uniform grid is shown as 10 \% contours in the background. (B) Absorptive 2DES spectra measured at a sample OD of 1.4 mOD reconstructed using a 15 $\times$ 15 ES sampled grid at pump-probe waiting time $T$ of 0 fs, 100 fs, and 1 ps. Contours are drawn at the 5$\%$, and 10-100$\%$ in 10$\%$  intervals for positive or negative contours.
	}
	\label{fig:fig7}
\end{figure*}

Comparison between SRPP and 2D experiments for the case of 1 mOD (Fig.~\ref{fig:fig6}) suggests that the AOPM approach to fPP for reconstructing $\omega_1$ or $\omega_3$ dimensions separately, with significantly faster data collection time may be an effective strategy to improve sensitivity for the lowest ODs. To further pursue this strategy, or when 2D CMs are desirable, reduction in ($t_{21},t_{43}$) grid is required even after optimizing $T$ sweep parameters. As an exploratory approach which rests on the flexibility of continuous $T$ scanning, we extend the idea of biased sampling of uniform signal contours in the ($t_{21},t_{43}$) time grid (Section \ref{collection}) to non-uniform sampling. Instead of sampling a uniform grid, we apply a biased exponential sampling (ES) scheme where the separation between the sampled data points increases exponentially with the delay. The ES points are derived using the equation $ P _n= \alpha *e^{\beta *n} - c$, where $P$ are new sampled points and $n$ is the index number correspond to each sampled point, and constants $\alpha = 1.9$, $\beta = 0.223$ and $c = 2$ are optimized for the particular system studied here. A schematic of this ES grid is shown in Fig.~\ref{fig:fig7}(A). For comparison, a 52 $\times$ 52 grid (with 1 fs time steps) is overlaid in the background. Such a grid is only sampled at $15 \times 15$ number of points in the ES scheme, which should result in a 91\% reduction in experimental data collection time. \\ 

Reconstruction of a uniform frequency grid starting from a non-uniformly sampled time-domain data is, in principle, guaranteed due to the linearity of Fourier transformation. Starting with $N$ ES points, $P_N$ in the time domain, one can write the Fourier transformation from $P_N$ to $F_K$ as $P_N = [iFT]_{N\times K} F_K$, where $F_K$ is defined in frequency domain and $[iFT]_{N\times K}$ is the inverse Fourier transformation matrix. Such a transformation can be found using compressive sensing (CS) algorithms by minimizing the $l_1$ norm of $F_K$. One such method, the spectral projected gradient approach, SPGL1\cite{spgl1}, was employed by Marcus and Aspuru-Guzik et al.\cite{Sanders2012} to reconstruct 2D spectra of rubidium vapor by collecting a dense grid and then non-uniformly sampling the grid to reconstruct the spectrum obtained with the full grid. A similar approach has been demonstrated by Pullerits et al.\cite{Wang2020}. \\

As a departure from previous approaches, we have used the ES scheme and collected the 15 $\times$ 15 grid from an \textit{independent} experiment conducted on a different day such that the uniform and non-uniform data points do not share correlations. In our experience, deriving data points from a denser grid tends to maintain correlations better, leading to a much improved reconstruction. However, such an approach does not translate to an actual reduction in experimental data collection time, which in going from 18$\times$18 to 15$\times$15 grid is nearly 1/3$^{rd}$ in our approach. The resulting absorptive 2D spectra, with both frequency axes reconstructed using SPGL1 algorithm applied on the 15$\times$15 ES grid, are shown in Fig.\ref{fig:fig7}(B). The sample OD was 1.4 mOD, with equivalent 2D spectra collected from a uniformly sampled 18$\times$18 grid shown in Fig.~\ref{fig:fig4}(bottom panel). 2D peak shifts of the order of $\omega_{1,3}$ resolution are evident, although peak positions and solvation dynamics features such as loss of frequency correlations, associated changes in 2D lineshapes and detection frequency are captured to within 10-20\% contour level. This is significant in light of the 1/3$^{rd}$ reduction in data collection time for starting sample OD ~2-3 orders of magnitude lesser than typical. 

The ES sampling scheme is made possible because of a combination of physical undersampling in AOPM approach and the flexibility of choosing ($t_{21},t_{43}$) grid points in the continuous $T$ scanning approach. Compared to several other implementations\cite{OgilvieARPC} of conventional 2DES, fMES is a slow time-domain experiment due to the requirement of sampling all the three time delays. Although CS applications in fMES have been quite sparse\cite{Sanders2012,Wang2020}, our proof-of-concept demonstration suggests that this could be an interesting avenue to explore further.

\section{Conclusions}
We have presented a visible WLC based fMES spectrometer that combines the advantages of physical undersampling and phase-sensitive lock-in detection in the AOPM approach with rapid scanning of the pump-probe delay $T$. Absorptive 2DES spectra and associated polar solvation dynamics features are consistent across a range of $~\sim$1-100 mOD, where the lowest sample ODs are $\sim$2-3 orders of magnitude lower than that reported in convention 2DES approaches. As a measure of sensitivity, suppression of 1/$f$ laser noise due to lock-in detection with a fast time-constant, and increased number of $T$ averages enabled by rapid scanning allow us to consistently measure coherent vibrational wavepackets even at the lowest sample ODs. This is especially significant for measurement of weak signals with dominant probe noise\cite{Moon1993}, where $\sim$5\% RMS spectral fluctuations across the entire visible WLC bandwidth are typical\cite{Ernsting2010,Bradler2014, Lang2018}. We have also demonstrated a significantly faster experiment with only detection frequency information but with features such as isolation of rephasing and non-rephasing pathways along with the quantum beat phase information not available in conventional pump-probe spectroscopy. A faster experiment with increased $T$ averaging can further improve the SNR at the lowest sample ODs, can provide excitation and detection frequency information separately, and still provide signal pathway isolation similar\cite{Senlik2015} to that available in a significantly slower 2D CM experiment. 

A continuous pump-probe delay scan per ($t_{21}, t_{43}$) point offers certain distinct advantages such as minimized sample exposure window relevant for spatially-resolved measurements on samples susceptible to photo-bleaching, fine $T$ sampling for effective averaging of high-frequency noise in the demodulated signal, such as sample scatter in the case of conventional 2DES on photosynthetic cells\cite{Dahlberg2013, Dahlberg2017}, and the ability to choose ($t_{21}, t_{43}$) time points. The latter feature leverages physical undersampling possible in the AOPM approach, and has also allowed us to explore further reductions in experimental time. An exponential sampling scheme biased towards ($t_{21}, t_{43}$) points where the signal is maximum allows 1/3$^{rd}$ reduction in data collection time. 2DES spectra reconstructed through compressive sensing algorithm are fairly consistent with uniformly sampled spectra to within $\sim$10-20\% contour motivating further applications of CS in fMES. Note that such a reduction in $T$ scan time is also dependent on the vibrational frequency being sampled, with further room available for faster $T$ scan velocity.  Recent demonstration\cite{Li2021} of fMES for detection of phonon wavepackets in single-layer MoSe$_2$ at room temperature have been quite promising. The advantages of rapid $T$ scanning approach for high-sensitivity spatially-resolved measurements, at lower repetition rates and minimized sample exposure, suggests promising future experiments.

\section*{Supplementary Material}
See supplementary material for details of spot size determination at sample position, simulation of the two photon autocorrelation signal using experimental laser spectrum, sample preparation, sample stability over experimental time, pump-probe delay zero determination and its reproducibilty check, and phasing of 2D spectra.

\section*{Acknowledgments}
AS and SP acknowledge the research fellowship from Indian Institute of Science(IISc). VNB acknowledge the Inspire fellowship from Department of Science and Technology. VT acknowledges support from Infosys Foundation Young Investigator Award. This project is supported in parts by Science and Engineering Research Board, India under grant sanction numbers CRG/2019/003691, Department of Atomic Energy, India under grant sanction number 58/20/31/2019-BRNS, Department of Biotechnology, India under grant sanction number BT/PR38464/BRB/10/1893/2020 and ISRO-STC grant sanction number ISTC/CSS/VT/468.

\section*{Data Availability}
The data that support the findings of this study are available from the corresponding author upon reasonable request

\bibliographystyle{unsrt}
\bibliography{OxzSolvation}

\end{document}